\documentclass[10pt,a4paper]{article}
\pdfoutput=1

\usepackage[utf8]{inputenc}
\usepackage[T1]{fontenc}
\usepackage{graphicx}
\usepackage{amscd}
\usepackage{amssymb}
\usepackage{amstext}
\usepackage{amsmath}
\usepackage{amsthm}
\usepackage{array}
\usepackage{booktabs}
\usepackage{hyperref}
\usepackage{listings}
\usepackage{xcolor}
\colorlet{Black}{black}

\usepackage[normalem]{ulem}
\usepackage{upgreek}
\usepackage{algorithm}{}
\usepackage{makecell}

\usepackage{tabularx}
\newcolumntype{C}{>{\centering\arraybackslash}X}

\usepackage{tikz}
\usetikzlibrary{arrows,shapes,patterns,decorations.text,decorations.pathreplacing}
\usetikzlibrary{automata,positioning}
\usetikzlibrary{trees}
\usepackage{pgfplots}
\usetikzlibrary{pgfplots.groupplots}

\usetikzlibrary{trees}
\usepackage[
    backend=bibtex,
    style=numeric,
    maxbibnames=9
]{biblatex}
\newtheorem*{remark*}{Remark}

\usepackage[nochapters]{classicthesis}
\usepackage[euler-digits,euler-hat-accent]{eulervm}
\usepackage{ stmaryrd }
\usepackage{ amssymb }

\DeclareMathOperator{\eps}{\emph{skip}}
\DeclareMathOperator{\outs}{\emph{out}}
\DeclareMathOperator{\trigs}{\emph{trig}}

\title{Sequential Circuits\\from Regular Expressions\\Revisited}
\author{Dogan Ulus}
\date{\today}
\begin{document}

\maketitle 

\begin{abstract}
We revisit the long-neglected problem of sequential circuit constructions from regular expressions. 
The class of languages that are recognized by sequential circuits is equivalent to the class of regular languages. 
This fact is shown in~\cite{copi1958realization} together with an inductive construction technique from regular expressions.
In this note, we present an alternative algorithm, called the trigger-set approach, obtained by reversing well-known follow-set approach to construct automata.
We use our algorithm to obtain a regular expression matcher based on sequential circuits. 
Finally, we report our performance results in comparison with existing automata-based matchers.
\end{abstract}

\section{Introduction}
A sequential circuit receives an input value at each step, updates its (internal) state, and yields an output value.
The current state of a circuit represents the history of input values received during its operation.
Circuits store their current state inside memory elements (e.g. flip-flops or registers) and update it with respect to newly received input value.
Then, the circuit yields an output value at each step, which is determined by the current state and input in general.
Clearly this is a simple but powerful model of computation and apparently impressed many people from electrical engineering and computer science over years.
While engineers were building modern computers upon sequential circuits, computer scientists studied abstract models of them.
Today we call these abstract models \emph{finite automata} but their original relation to sequential circuits is largely forgotten.

In this note, we study sequential circuits with a goal to obtain regular expression matchers, which is an application historically that employs automata and related techniques~\cite{thompson,aho2007compilers}.
Interestingly enough, the use of automata were also forgotten for pattern matching purposes in the sake of search-based (backtracking) methods until it is revived in a blog post\footnote{\url{https://swtch.com/~rsc/regexp/regexp1.html}} by Russ Cox. 
Eventually this effort has provided us industrial-grade pattern matchers that use automata (again) such as Google's \textsc{re2} engine\footnote{\url{https://github.com/google/re2}}.
Besides recent programming languages such as \textsc{go} and \textsc{rust} implement automata inside their regular expression engines unlike \textsc{perl} and \textsc{python} that implement backtracking.
It is worth to note that existing backtracking and automata based matchers are mostly comparable in performance.
However, there are (pathological) cases that backtracking fails miserably and requires exponential time for matching whereas automata have a formal guarantee for linear execution time.

So what can we gain from sequential circuits in this already-quite-developed world of pattern matching?
First, we do not want a~worst-case exponential time algorithm, which can be more critical as we will eventually go beyond text processing as in~\cite{patterns,timed-deriv}.
Therefore, we eliminate backtracking matchers from our discussion and focus on the other class.
Automata based matchers usually implement two types\footnote{Some can have more. For example, \textsc{re2} implements an intermediate form called one-pass \textsc{nfa} for patterns having limited non-determinism.} of automata, namely deterministic (\textsc{dfa}s) and non-deterministic (\textsc{nfa}s).
Roughly speaking, for a given expression with a size $m$, \textsc{dfa}s tend to be large (exponentially bounded by $m$) and fast (constant to $m$) whereas \textsc{nfa}s are small in size (linearly bounded by $m$) and slow (quadratic to $m$).
Note that actual performances considerably depend on the expression as well as the input string to be matched.
Sequential circuits would reside at the non-deterministic side of this spectrum but they seem to be better to handle non-determinism than \textsc{nfa}s.
It is not wrong to consider sequential circuits to be a yet another way to implement non-deterministic automata ---remember that automata are abstract models of sequential circuits--- but we argue that this view is missing some essence.

There is one important operational contrast between automata and sequential circuits regarding their update mechanisms.
For automata, update mechanism operates by pushing the information forward. 
The information about being in a state is propagated to the next states with respect to the current input so we compute reachable successors. 
On the contrary, update mechanism for sequential circuits operates by pulling the information from back.
This time we independently compute the reachability of each state from predecessors with respect to the current input.
By analogy, we can say that automata run by a rear-wheel-drive whereas sequential circuits a front-wheel one.
Both computations are going to exactly the same direction but the difference in update mechanisms creates some interesting trade-offs in practice.
Maybe a \textsc{4}-wheel drive is what we want.

In the following, we first review sequential circuit briefly and then propose an algorithm to directly construct sequential circuits from (classical) regular expressions. 
The last section reports the implementation, test cases, and some comparison with \textsc{re2}'s \textsc{dfa} and \textsc{nfa} implementations.

\section{Sequential Circuits}
\begin{figure}[t]
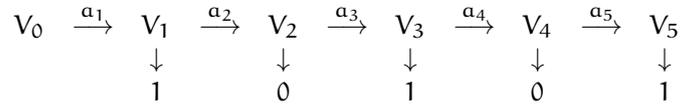

\begin{equation*}
\begin{array}{ccccccccccc}
V_0 & \overset{a_1}{\longrightarrow} & V_1 & \overset{a_2}{\longrightarrow} & V_2 & \overset{a_3}{\longrightarrow} & V_3 & \overset{a_4}{\longrightarrow} & V_4 & \overset{a_5}{\longrightarrow} & V_5\\
&  & \downarrow &  & \downarrow &  & \downarrow &  & \downarrow & & \downarrow \\
&  & 1 &  & 0 &  & 1 & & 0 & & 1 \\
\end{array}
\end{equation*}
\caption{A sequential circuit computation.}
\label{fig:sequential}  
\end{figure}

A sequential circuit reads a sequence $X_{1}X_{2}\dots$ of $k$-dimensional Boolean vectors such that $X_{n}\in\mathbb{B}^{k}$ for $n = 1, 2, \dots$.
Clearly any finite alphabet~$\Sigma$ can be encoded as a set of $k$-dimensional Boolean vectors for a suitable $k$ with typical examples of \textsc{ascii} and Unicode.
Therefore, in this note, we interchangeably use Boolean vectors and letters as inputs to circuits.
Then, we characterize a sequential circuit by three elements as follows.
\begin{equation*}
\begin{array}{rcl}
	V & : & \text{$m+1$ dimensional Boolean state vector}\\
	F_{i}(V, X) & : & \text{Next-state functions for each position $i = 0, 1,\dots, m$}\\
	Y(V, X) & : & \text{The output function of the circuit}
\end{array}
\end{equation*}
such that $X\in\mathbb{B}^{k}$ is the current input value.
The functions $F_{i}$ and $Y$ are Boolean functions from $\mathbb{B}^{m+1} \times \mathbb{B}^{k}$.
We denote by $V_{n}$ the valuation (content) of the vector $V$ at the time step $n$ and we call $V_{0}$ the initial valuation of the circuit.
The circuit operates by reading the current input value $X_{n}$, updating the state vector $V$ such that
\begin{equation}
	V_{n}(i) = F_{i}(V_{n-1}, X_n) \quad\text{for } i = 0, \dots, m
\end{equation}
and yielding a Boolean value $Y_{n}$ such that
\begin{equation}
	Y_{n} = Y(V_{n-1}, X_n)
\end{equation}
at each time step $n$. 
The output of a sequential circuit is completely determined by the initial valuation $V_{0}$ and the sequence of input values read so far.
This is called sequential model of computation, which relates input sequences to output sequences as illustrated in Figure~\ref{fig:sequential}.
It is customary to say that any sequence of input values that make the circuit to yield $1$ is \emph{accepted} by the circuit for a fixed initial valuation.
For example, input sequences $a_1$, $a_1a_2a_3$, and $a_1a_2a_3a_4a_5$ are accepted by the circuit in Figure~\ref{fig:sequential} whereas $a_1a_2$ and $a_1a_2a_3a_4$ are rejected.
The set of all input sequences accepted by a sequential circuit is called the language of the circuit.
It is shown in~\cite{copi1958realization} that all and only regular languages recognized by sequential circuits as an analogue of Kleene's theorem~\cite{kleene}.

These concepts are of course familiar to anyone who studied automata theory, which is established in classical papers~\cite{mealy1955method,moore1956gedanken,rabinscott} as an abstract treatment of sequential circuits.
%
%\footnote{Do not confuse the term sequential here with the one as in sequential/parallel programming. The former emphasizes sequentiality in time whereas the latter sequentiality of . Surely sequential computations can be executed in parallel.}.
%
However, reinstalled circuit approach here makes it clearly visible that checking acceptance or matching a word can be reduced to a computation of a set of Boolean functions repeatedly. 
Especially we emphasize that (more) expensive list processing operations in \textsc{nfa} simulations can be avoided if we had constructed circuits instead of automata.
Besides the efficiency can be easily increased by exploiting parallelism of the hardware if needed. 
On a fully parallel hardware\footnote{We say so because our circuit definition is still more mathematical than electrical.}, circuits are probably the most efficient representations of regular languages since they would have a small (linear) size and operate very fast (one clock period per one letter of the input).
This fact is rather well-known in hardware communities~\cite{sidhu2001fast,hardware-assertions} but methods involve a translation into \textsc{dfa} or \textsc{nfa}, which brings additional work, if not complexity (e.g. $\epsilon$-transitions).

\section{Sequential Circuits from Regular Expressions}

In this section, we present an algorithm to construct sequential circuits directly from regular expressions, which are syntactic representations of regular languages. 
We give the syntax of regular expressions in this note by the following grammar.
\begin{equation*}
E :=  a\ |\ E_{1} \cup E_{2}\ |\ E_{1} \cdot E_{2}\ |\ E^{*}\ |\ E^{+}\ |\ E^{?}
\end{equation*}
where $a$ is a letter of an alphabet $\Sigma$ and $E$ is a regular expression. The operators $\cdot$ and $\cup$ denote concatenation and union as usual. Other operators $*$, $+$, and $?$ are repetition operators and sometimes called zero-or-more, one-or-more, and zero-or-one repetition, respectively.

Our algorithm falls into the family of position-based construction algorithms that associate each letter (or sub-expression) of the expression with a unique position value. 
Previous works~\cite{mcnaug-yamada,glushkov,aho2007compilers,cont-berry-sethi} in this family\footnote{Inductive constructions such as Thompson's~\cite{thompson} are also in this family as positions are implicitly assumed. The other family is called derivative-based constructions stemmed from Brzozowski's derivatives~\cite{brzo}. Berry and Sethi show that these two happy families are alike~\cite{cont-berry-sethi}.} are interested in computing succeeding positions (follow sets) for each position in the expression.
The rough idea here is that the acceptance procedure for an input word would resemble the act of going from position to position on the expression.
Consequently, these positions become states of an \textsc{nfa} and follow sets make the transition function.
On the other hand, we are interested in computing trigger conditions (or preceding positions) for each position to obtain a sequential circuit.
To this end, we closely follow the work of Berry and Sethi~\cite{cont-berry-sethi} but reverse the follow set approach. 
The proposed treatment is mostly symmetric and yields a sequential circuit instead of an \textsc{nfa} as the outcome.
This result can be taken as yet another demonstration of the close relation between automata and sequential circuits.

We start by marking mark all letters of a regular expression $E$ to make them distinct. 
The marked version of a regular expression $E$ is obtained by associating letters in $E$ with a number called position. 
Positions are denoted by subscripts over letters. 
In the strict sense, marked letters $a_{i}$ and $a_{j}$ are distinct if $i \neq j$.
We designate the position $0$ as the initial position of the expression and then enumerate letters left to right. 
The leftmost letter in $E$ is associated with the position $1$ and the rest goes on. 
For example, an expression $(a\cdot b \cup b)^{*}\cdot b \cdot a$ becomes $(a_1\cdot b_2 \cup b_3)^{*}\cdot b_4 \cdot a_{5}$ after marking process.
Finally, we say that the size $\#(E)$ of an expression $E$ is the number of letters in the expression.

In the following, we present three functions, $\eps$, $\outs$, and $\trigs$ that have been used to construct a sequential circuit from a marked regular expression.
%
% All three functions work on the syntax tree of the expression.
%
First, the function $\eps$ checks whether the language of an expression $E$ contains the empty word (thus whether $E$ is skippable). 
This function can be computed inductively for regular expressions by the following rules.
\begin{equation*}
  \begin{array}{rcl}
    \eps(a_i) & = & \textsc{0} \\ 
    \eps(E_{1}\cup E_{2}) & = & \eps(E_{1})\text{ or }\eps(E_{2}) \\
    \eps(E_{1}\cdot E_{2}) & = & \eps(E_{1})\text{ and }\eps(E_{2})\\
    \eps(\varphi^{*}) & = & \textsc{1}\\
    \eps(\varphi^{+}) & = & \eps(\varphi)\\
    \eps(\varphi^{?}) & = & \textsc{1}\\ 
  \end{array}
\end{equation*}
Second, we define the $\outs$ function that computes the outputting (or last) positions of each node. 
The computation of the $\outs$ function is inductively given by the following rules.
\begin{equation*}
  \begin{array}{rcl}
    \outs(a_i) & = & \{i\} \\ 
    \outs(E_{1}\cup E_{2}) & = & \outs(E_{1})\cup\outs(E_{2}) \\
    \outs(E_{1}\cdot E_{2}) & = & \begin{cases}
    \outs(E_{1})\cup\outs(E_{2})& \text{if }\eps(E_2)\\
    \outs(E_{2})&\text{otherwise.}
    \end{cases}\\
    \outs(E^{*}) & = & \outs(E)\\
    \outs(E^{+}) & = & \outs(E)\\
    \outs(E^{?}) & = & \outs(E)\\ 
  \end{array}
\end{equation*}
\begin{figure}[b]
\begin{tikzpicture}[
  tlabel/.style={pos=0.4,right=-1pt,font=\footnotesize\color{red!70!black}}, grow=left, node distance=0.1cm
]
\node (root) {$\cdot$}
child {node (root1) {$\cdot$}
 	child{node[level 1/.style={sibling distance=40mm}, level 2/.style={sibling distance=20mm}] (root2) {$^{*}$}
  	  child{node (root3) {$\cup$}
  		child{node (root4) {$\cdot$}
  		  child{node (1) {$a_{1}$}}
  		  child{node (2) {$b_{2}$}}
  		}
  		child{node (3) {$b_{3}$}}	
  	  }
  	}
  child{node (4) {$b_{4}$}}
}
child {node (5) {$a_5$}};

\node[left=of 1] (s1) {\makecell[r]{$\eps = 0$\\$\outs = \{1\} $}};
\node[left=of 2] (s2) {\makecell[r]{$\eps = 0$\\$\outs = \{2\} $}};
\node[below=of 3] (s3) {\makecell[r]{$\eps = 0$\\$\outs = \{3\} $}};
\node[below=of 4] (s4) {\makecell[r]{$\eps = 0$\\$\outs = \{4\} $}};
\node[below=of 5] (s5) {\makecell[r]{$\eps = 0$\\$\outs = \{5\} $}};

\node[above=of root] (s3) {\makecell[r]{$\eps = 0$\\$\outs = \{5\} $}};
\node[above=of root1] (s3) {\makecell[r]{$\eps = 0$\\$\quad\quad\outs = \{4\} $}};
\node[above=of root2] (s3) {\makecell[r]{$\eps = 1$\\$\quad\quad\outs = \{2,3\} $}};
\node[above=of root3] (s3) {\makecell[r]{$\eps = 0$\\$\outs = \{2,3\} $}};
\node[above=of root4] (s3) {\makecell[r]{$\eps = 0$\\$\outs = \{2\} $}};

\end{tikzpicture}
\caption{Computing $\eps$ and $\outs$ over the syntax tree of {\small$\footnotesize(a_1\cdot b_2 \cup b_3)^{*}\cdot b_4 \cdot a_{5}$}.}
\label{fig:tree}   
\end{figure}
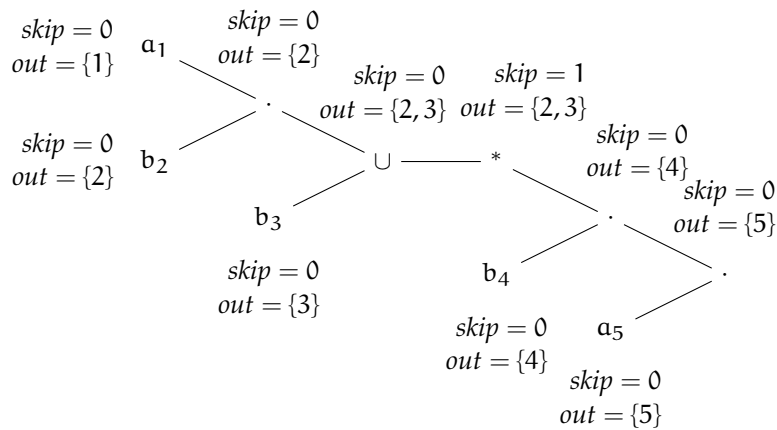

Continuing our running example, we compute $\eps$ and $\outs$ functions and illustrate over the syntax tree in Figure~\ref{fig:tree}.
We annotate each node of the syntax tree with corresponding value computed by $\eps$ and $\outs$ functions.

Third, recall that we have reserved the position $0$ as the starting position.
We use the position $0$ as the trigger for the expression and it allows us to control where to start a match on the input sequence.
We will explain this point later with an example.  
The function $\trigs(E, \{0\})$ yields a set of triples $(i, a, H)$ for a given regular expression $E$ where $i$ is a position, $a$ is the corresponding letter, and $H$ is a set of positions that trigger the position $i$ together with $a$. 
In other words, one should be in a position $j\in H$ and read the letter $a$ to reach the position $i$ in the next step. 
Then, we inductively define the function $\trigs$ as follows.

\begin{equation*}
  \begin{array}{rcl}
    
    \trigs(a_i, H) & = & \{(i, a, H)\} \\ 
    \trigs(E_{1}\cup E_{2}, H) & = & \trigs(E_{1}, H)\cup\trigs(E_{2}, H) \\
    \trigs(E_{1}\cdot E_{2}, H) & = & \begin{cases}
    \trigs(E_{1}, H)\cup\trigs(E_{2},\ \outs(E_1) \cup H) & \text{if }\eps(E_1)\\
    \trigs(E_{1}, H)\cup\trigs(E_{2},\ \outs(E_1))&\text{otherwise.}
    \end{cases}\\
    \trigs(E^{*}, H) & = & \trigs(E,\ \outs(E) \cup H)\\
    \trigs(E^{+}, H) & = & \trigs(E,\ \outs(E) \cup H)\\
    \trigs(E^{?}, H) & = & \trigs(E, H)\\ 
  \end{array}
\end{equation*}

An example computation of trigger sets is give over our running example in Figure~\ref{fig:trig}.
By computing $\trigs(E, \{0\})$, we have almost finished our sequential circuit construction but we must also define initializations and the output function. 
The output function is determined by $\outs(E)$ function at the top level. 
For the starting behavior, we have two typical options: (1) Start matching only from the beginning of the input sequence. This is also known as the acceptance, full match, etc. (2) Start anywhere on the input sequence. This is also known as the suffix acceptance, partial match, etc., which is indeed equivalent to the acceptance problem of $\Sigma^{*}\cdot E$.
These starting behaviors is determined by the next-state function $F_{0}$ of the initial position.
Finally, we present the full construction in Algorithm~\ref{algo:re2seq} and the circuit generated for the expression $\big((a\cdot b) \cup b \big)^{*}\cdot b \cdot a$ in Figure~\ref{fig:circuit}.

\begin{figure}[t]
\begin{center}
\begin{tabular}{ccc}
\toprule
Position & Letter & Trigger Set\\
\midrule
$1$ & $a$ & $\{ 0, 2, 3\}$\\
$2$ & $b$ & $\{ 1\}$\\
$3$ & $b$ & $\{ 0, 2, 3\}$\\
$4$ & $b$ & $\{ 0, 2, 3\}$\\
$5$ & $a$ & $\{ 4 \}$\\
\bottomrule
\end{tabular}
\end{center}
\caption{Trigger sets computed by $\trigs(E, \{0\})$ where {\small$E = (a_1\cdot b_2 \cup b_3)^{*}\cdot b_4 \cdot a_{5}$}.}
\label{fig:trig}   
\end{figure}

\begin{algorithm}[h!]
\caption{Trigger Set Algorithm}
\label{algo:re2seq}
For a given regular expression $E$, a sequential circuit $(V, F, Y)$ is obtained as follows.
\begin{enumerate}
  \item Compute $\#(E)$, $\eps(E)$, $\outs(E)$, and $\trigs(E, \{0\})$ functions.
  \item Allocate an $m$+1 dimensional state vector $V$ and initialize $V$ such that $V_{0}(0):=1,\ V_{0}(i):=0$ for $i = 1,\dots,m$ where $m = \#(E)$.
  \item For the initial position, define either
  \begin{enumerate}
    \item $F_{0} := 0$ to start matching only from the first letter, or
    \item $F_{0} := 1$ to start matching anywhere on the input sequence.
  \end{enumerate} 

  \item For each triple $(i, a, H)\in\trigs(E)$, define the rest of next state functions
    $$F_{i} :=  (X = a)\ \wedge\ \bigvee_{j\in H} V(j)$$
    where $X$ is the current input value.

  \item Define the output function $Y := \bigvee_{i\in\outs(E)} F_i$.
\end{enumerate}
\end{algorithm}

\begin{figure}[t]
\centering
\begin{equation*}
\begin{array}{rcl}
    V_0 & := & (1,0,0,0,0,0)\\
    \\
    F_{0} & := &  0\\
    F_{1} & := &  (X = a)\ \wedge\ ( V(0) \vee V(2) \vee V(3))\\
    F_{2} & := &  (X = b)\ \wedge\ V(1)\\
    F_{3} & := &  (X = b)\ \wedge\ ( V(0) \vee V(2) \vee V(3))\\
    F_{4} & := &  (X = b)\ \wedge\ ( V(0) \vee V(2) \vee V(3))\\
    F_{5} & := &  (X = a)\ \wedge\ V(4)\\
    \\
    Y & := & (X = a)\ \wedge\ V(4)
\end{array}
\end{equation*}
\caption{A sequential circuit that recognize the expression $\big((a\cdot b) \cup b \big)^{*}\cdot b \cdot a$ constructed by trigger set algorithm.}
\label{fig:circuit}   
\end{figure}

\begin{figure}[b!]
\begin{lstlisting}[frame=single, language=c, breaklines=true, floatplacement=t, basicstyle=\ttfamily, tabsize=2, caption=Code generated for \texttt{(((a;b)|b)*);b;a}]
#include <iostream>
#include <fstream>
#include <cstring>

int main(int argc, char **argv) {
	int state[6] = {1,0,0,0,0,0};
	int next_state[6] = {0,0,0,0,0,0};
	std::ifstream ifs(argv[1]);
	std::string word((std::istreambuf_iterator<char>(ifs)),(std::istreambuf_iterator<char>()));
	for (char letter : word){
		next_state[0] = 1; // Start anywhere
		next_state[1] = (letter == 'a') and (state[0] or state[2] or state[3]);
		next_state[2] = (letter == 'b') and (state[1]);
		next_state[3] = (letter == 'b') and (state[0] or state[2] or state[3]);
		next_state[4] = (letter == 'b') and (state[0] or state[2] or state[3]);
		next_state[5] = (letter == 'a') and (state[4]);
		std::memcpy(state, next_state, sizeof(state));
	}
	std::cout << next_state[5] << std::endl;
}
\end{lstlisting}
\label{fig:code}
\end{figure}

\section{Implementation}

The prototype implementation\footnote{\url{https://github.com/doganulus/reelay}} of the trigger set algorithm performs two main tasks. 
The first task is in computing $\#$, $\eps$, $\outs$, and $\trigs$ functions for the regular expression given.
Corresponding visitor implementations annotate the syntax tree according to definitions in the previous section and return the size of the expression, the output set, and trigger sets for each position.
Then, using this information, we construct a sequential circuit from the expression, that is to say, generate a \textsc{c++} code to be compiled into a program that matches the expression over the input word.
Generated code is very simple and currently requires the name of a file that contains the input word as its only argument.
For example, in Listing 1, we show the code generated for the sequential circuit given in Figure~\ref{fig:circuit}.

\begin{table}[b!]
\caption{\small$\big((a\cdot b) \cup b \big)^{*}\cdot b \cdot a$ over $\Sigma = \{a, \dots, z\}$}
\begin{tabularx}{\textwidth}{CCC}
\toprule
\textsc{dfa} & \textsc{nfa} & Sequential \\
\midrule
152 \textsc{mb}\textbackslash s & 13 \textsc{mb}\textbackslash s& \textbf{268} \textsc{mb}\textbackslash s\\
\bottomrule
\end{tabularx}  
\caption{\small$a\cdot b\cdot c\cdot d\cdot e\cdot f\cdot g\cdot h\cdot i\cdot j\cdot k\cdot l\cdot m\cdot n\cdot o\cdot p\cdot q\cdot r\cdot s\cdot t\cdot u\cdot v\cdot w\cdot x\cdot y\cdot z$}
\begin{tabularx}{\textwidth}{CCC}
\toprule
\textsc{dfa} & One-pass \textsc{nfa}? & Sequential \\
\midrule
\textbf{304}\textsc{mb}\textbackslash s& 161 \textsc{mb}\textbackslash s& 48\textsc{mb}\textbackslash s\\
\bottomrule
\end{tabularx}  
\caption{\small$(x\cup y\cup z)\cdot a\cdot b\cdot c\cdot d\cdot e\cdot f\cdot g\cdot h\cdot i\cdot j\cdot k\cdot l\cdot m\cdot n\cdot o\cdot p\cdot q\cdot r\cdot s\cdot t\cdot u\cdot v\cdot w\cdot x\cdot y\cdot z$}
\begin{tabularx}{\textwidth}{CCC}
\toprule
\textsc{dfa} & \textsc{nfa} & Sequential \\
\midrule
\textbf{135}\textsc{mb}\textbackslash s& 30\textsc{mb}\textbackslash s& 44\textsc{mb}\textbackslash s\\
\bottomrule
\end{tabularx}
\caption{\small$(a^{?})^{n}\cdot a^{n}$ over $\Sigma = \{a, \dots, z\}$}
\begin{tabularx}{\textwidth}{cCCC}
\toprule
n & \textsc{dfa} & \textsc{nfa} & Sequential \\
\midrule
10 & \textbf{335}\textsc{mb}\textbackslash s& 38\textsc{mb}\textbackslash s& 257\textsc{mb}\textbackslash s\\
20 & \textbf{335}\textsc{mb}\textbackslash s& 22\textsc{mb}\textbackslash s& 170\textsc{mb}\textbackslash s\\
30 & \textbf{335}\textsc{mb}\textbackslash s& 15\textsc{mb}\textbackslash s& 115\textsc{mb}\textbackslash s\\
\bottomrule
\end{tabularx}
\caption{\small$((a \cup b)^{*})\cdot a \cdot (a \cup b)^{n}$ over $\Sigma = \{a,b\}$}
\begin{tabularx}{\textwidth}{cCCC}
\toprule
n & \textsc{dfa} & \textsc{nfa} & Sequential \\
\midrule
10 & \textbf{93}\textsc{mb}\textbackslash s& 5.4\textsc{mb}\textbackslash s& 35\textsc{mb}\textbackslash s\\
14 & \textbf{37}\textsc{mb}\textbackslash s& 4.3\textsc{mb}\textbackslash s& 19\textsc{mb}\textbackslash s\\
15 & 4.3\textsc{mb}\textbackslash s& 4.3\textsc{mb}\textbackslash s& \textbf{17}\textsc{mb}\textbackslash s\\
20 & 3.3\textsc{mb}\textbackslash s& 3.3\textsc{mb}\textbackslash s& \textbf{12}\textsc{mb}\textbackslash s\\
30 & 2.7\textsc{mb}\textbackslash s& 2.7\textsc{mb}\textbackslash s& \textbf{7.2}\textsc{mb}\textbackslash s\\
\bottomrule
\end{tabularx}
\end{table}

We compare our sequential circuit based implementation against automata based regular expression matcher \textsc{re2}'s \textsc{dfa} and \textsc{nfa} implementations.
We generate a similar code that uses \textsc{re2} for testing purposes. 
To this end, we simply call \textsc{re2}'s corresponding matching function (\texttt{RE2::PartialMatch}) instead of our sequential circuit implementation.
Finally note that \textsc{re2} constructs a \textsc{dfa} by default. 
In order to force \textsc{re2} to construct \textsc{nfa}, we reduce maximum allowed cache for \textsc{dfa} simulation so that it falls back to \textsc{nfa}.
These codes that use \textsc{re2} can be found in the appendix.

We compile the generated code using standard \texttt{g++} compiler with level two optimizations (\texttt{-O2}).
In our tests, we perform suffix matching for a test pattern over very long (67M characters = 67\textsc{mb}) random sequences of textual characters over an alphabet $\Sigma$.
All tests run on a 3.3GHz machine.
Since the execution time is linear to the input size, we only report the throughput of matching (in \textsc{mb} per second) obtained by dividing input file size by the minimum of 10 actual execution times.
Then we present our experimental results in Tables 1-5 and discuss them in the following section. 

\section{Discussion}
In this note, we presented an algorithm to construct a sequential circuit from regular expressions.
We implemented our algorithm straightforwardly and tested sequential circuit approach against a well-engineered regular expression engine that implements automata.
In our tests, sequential circuits clearly outperforms standard \textsc{nfa}s, which shares similar memory requirements.
There is one case (shown in Table 2) that \textsc{nfa} seems faster but this is probably due to that \textsc{re2}'s engine falls back to one-pass \textsc{nfa} rather than standard \textsc{nfa} since the pattern is unambiguous.
It is no surprise that deterministic automata are faster than sequential circuits when the pattern is more deterministic and not too large.
However, there are patterns when the number of deterministic states grows exponentially in the size of expression.
The pattern given in Table 5 is not randomly selected and similar to the example used in~\cite{moore1971bounds}, which proves the exponential bound on \textsc{nfa}-\textsc{dfa} conversion is indeed tight.
In short, this is very bad news for \textsc{dfa}.
The \textsc{dfa} implementation in \textsc{re2} performs a lazy determinization procedure, which computes and stores new deterministic states only when it is required.
The number of deterministic states kept in the cache is capped by a fixed value (the default must be around 10K states) and the implementation falls back to \textsc{nfa} if this limit is violated often.
This perhaps explains the dramatic performance decrease observed for \textsc{dfa} between $n=14$ and $n=15$ in Table 5. 
On the contrary, sequential circuits perform better for such highly non-deterministic patterns.
Therefore, we believe that sequential circuits offer an alternative solution for the pattern matching problem and can replace non-deterministic automata to this end. 
%
% They still have some room for optimization and perhaps parallelism.

\printbibliography 

\clearpage
\appendix

\section{Generated Codes}

\begin{lstlisting}[frame=single, language=c++, breaklines=true, basicstyle=\ttfamily, caption=Example test code for RE2 (DFA)]
#include <iostream>
#include <fstream>
#include <re2/re2.h>

int main(int argc, char **argv) {

	std::ifstream ifs(argv[1]);
	std::string word((std::istreambuf_iterator<char>(ifs)),(std::istreambuf_iterator<char>()));

	    std::cout << RE2::PartialMatch(word, "(?:(?:(?:ab)|b)*)ba$") << std::endl;

}
\end{lstlisting} 

\begin{lstlisting}[frame=single, language=c++, breaklines=true, basicstyle=\ttfamily, caption=Example test code for RE2 (Forced NFA)]
#include <iostream>
#include <fstream>
#include <re2/re2.h>

int main(int argc, char **argv) {

    std::ifstream ifs(argv[1]);
    std::string word((std::istreambuf_iterator<char>(ifs)),(std::istreambuf_iterator<char>()));

    RE2::Options opt;
    opt.set_max_mem(2048);
    RE2 re("(?:(?:(?:ab)|b)*)ba$", opt);

    std::cout << RE2::PartialMatch(word, re) << std::endl;

}
\end{lstlisting} 
\end{document}